# Experiences and Insights from Applying GQM⁺Strategies in a Systems Product Development Organisation

Jürgen Münch*, Fabian Fagerholm*, Petri Kettunen*, Max Pagels*, Jari Partanen†

*University of Helsinki
Department of Computer Science
P.O. Box 68, FI-00014, FINLAND
Email: {firstname.lastname}@cs.helsinki.fi

†Elektrobit Corporation
Wireless Business Segment
Tutkijantie 8, FIN-90690, Oulu, FINLAND
Email: jari.partanen@elektrobit.com

*Abstract*—Aligning software-related activities with corporate strategies and goals is increasingly important for several reasons such as increasing the customer satisfaction in software-based products and services. Several approaches have been proposed to create such an alignment. GQM⁺Strategies is an approach that applies measurement principles to link goals and strategies on different levels of an organisation. In this paper, we describe experiences from applying GQM⁺Strategies to elicit, link, and align the goals of an integrated systems product development organisation across multiple organisational levels. We provide insights into how GQM⁺Strategies was applied during a five-month period. The paper presents the enacted application process and main lessons learnt. In addition, related approaches are described and an outlook on future work is given.

## I. Introduction

Many studies have shown that teams are effective if they have shared goals. There are also indications that organisations are "high-performing" if they are goal-oriented by nature [1]. This implies that organisations have a common understanding of their goals on different levels and that these goals are linked to each other. In practice, it is often challenging to evolve an organisation towards such a situation. In addition, identifying and setting the "right" goals is a complex process in itself and can be done in different ways, such as in a "push" or "pull" fashion. In order to ascertain whether a goal has been met, it should be coupled with appropriate metrics [2].

Given the hierarchical structure of most organisations, goals on lower organisational levels may differ from those on higher levels. Despite potential disconnects, goals on different organisational levels should be related to one another: a low-level goal that does not contribute towards the achievement of a higher-level goal may be indicative of an organisation that has become disintegrated or, worse yet, derailed.

GQM⁺Strategies is an approach designed to help software businesses define goals on different organisational levels and to model how these goals contribute to one another [3]. In this paper, we describe experiences from a case study in which GQM⁺Strategies was used to identify, link, and align the goals of an industrial integrated systems product development (IPD) organisation. The organisation under study develops integrated hardware/software platform products.

The rest of this paper is organised as follows. Section II describes the GQM⁺Strategies approach. Section III describes previous work on organisational visibility, alignment, and application of GQM⁺Strategies. Section IV discusses the research methodology and execution of this case study. Section V details the main findings of applying GQM⁺Strategies. Finally, Section VI shortly summarises the lessons learnt and outlines possible future work.

## II. GQM⁺Strategies

GQM⁺Strategies [3] is an approach that directs goals and strategies of an organisation towards value-creation. It aligns and integrates goals, strategies, and measurements at all levels of an organisation and thereby helps to identify misalignments. In addition, GQM⁺Strategies supports the clarification, harmonisation, and communication of goals and strategies by making them explicit and integrating them. It also facilitates the identification of emerging strategies in an organisation and helps to integrate them into an overall business strategy. From the operational point of view, measurement is used to define success criteria, to monitor the achievement of goals, and to analyse business performance for making informed decisions. The measurement principles are based on the Goal/Question/Metric paradigm (GQM) [4]. GQM⁺Strategies is domain-independent. Though initially developed for use in software engineering, it is not necessarily limited to this domain.

GQM⁺Strategies models goals, which specify the "what" of the organisational strategy, and associated strategies, which specify the "how". Figure 1 provides an overview of the concepts of GQM+Strategies. Goals can be defined on all levels of an organisation from the business level down to the project level and even lower. A strategy defines how one or more goals can be achieved. A strategy is usually associated with a set of concrete activities such as business processes and might be refined into lower-level goals. Typically, strategies are selected based on assumptions and known facts (i.e., context). An assumption could be, for instance, that the customer satisfaction with a specific product can be increased by decreasing the amount of defects that slip through the acceptance test. Such assumptions can be seen as hypotheses. Assumptions are typically used in GQM+Strategies to describe the relationship between a strategy and a goal or to describe the rationales that led to the definition of a specific goal.

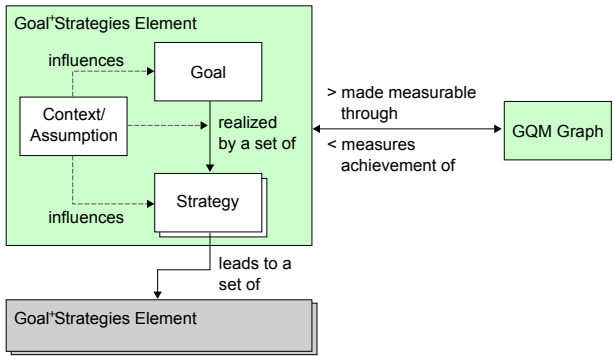

Fig. 1. GQM⁺Strategies Components (based on Basili et al. [3]).

After identifying a goal for a given organisational level (or unit), a potential strategy or set of strategies for fulfilling that goal is typically devised based on elicited assumptions and overall context. If it is possible to further refine the devised strategy into more tangible actions on lower organisational levels, a set of sub-goals is introduced, each with their own set of sub-strategies. Collectively, these sub-strategies contribute towards the achievement of their parent goal. The strategies are then implemented and their effects measured using GQM. The result of applying the GQM⁺Strategies approach is a so-called grid, i.e., a model that links goals and associated strategies for each level as well as associated facts, assumptions, and measures. GQM⁺Strategies also provides mechanisms for analysis and interpretation of measured data.

Although a top-down, iterative goal derivation process is typical, GQM⁺Strategies may be applied from any organisational level, moving either up or down [5]. Goals on high levels tend to be more general and abstract in nature, as do their immediate strategies. The corresponding pairs on lower organisational levels are typically narrower in scope, but more accurate in nature: strategies become more specific and concrete, with less room and need for interpretation. In practice, developing a detailed GQM⁺Strategies model is an inherently iterative process, ideally with frequent communication between the practitioners developing the model and representatives of the target organisation. In general, it is to be expected that a model will undergo significant changes in the beginning of its development process, and a period of relative stability towards the end. However, as nowadays the business and technical environment change at a high speed, a continuous review and evolution of the model is necessary. This includes a continuous validation of the underlying assumptions. Therefore, it is important to see the resulting model as a living artefact, which should be kept up-to-date when goal changes are needed, or the contextual factors or assumptions are revisited. By doing so, GQM⁺Strategies helps to avoid future misalignment of goals. Furthermore, when strategies are implemented with associated measurements, organisational measurement programs can be devised according to the explicit data gathering requirements, avoiding purposeless measurement efforts.

## III. BACKGROUND AND RELATED WORK

In this chapter, we describe work related to experiences with organisation alignment and GQM⁺Strategies in particular. We mainly address the software development domain, although important approaches from the business domain and integrated product development systems engineering are also considered.

### A. Organisational Visibility and Alignment

High-performing R&D organisations should share information efficiently inside the organisation and between their customers and other stakeholders. Such transparency makes it possible to perform effectively and efficiently when all parties understand the common targets and align their efforts accordingly [6], [7].

Comprehensive visibility and alignment are of particular importance in integrated product development organisations [8], [9]. However, R&D units have inherent additional difficulties in achieving such visibility [10]: development projects are multidisciplinary by nature, requiring complicated – often virtual or even global – organisational structures and coordination of cross-functional teams over a product's entire life-cycle [11], [12], [13]. Furthermore, many advanced products are nowadays increasingly complex with various new technologies and multiple component sources [14]. Such an increase in complexity calls for more systematic and rigorous goal definition and management. Systems engineering and systems thinking in general become core competencies and competitive advantages [15], [16], [17].

Product creation platforms are particular forms of integrated product development services. They are becoming more commonplace in modern product development ecosystems where multiple different vendors collaborate in flexible networked settings for value creation [18]. In such environments, there is a need for consistent project and product management. Not only are product structures more complex, there are also multiple stakeholders spanning organisational boundaries – e.g., component suppliers, customisation specialists, platform customers, end-users. Often, they have to deal with building or participating in whole ecosystems. Shared, transparent goals and their proper alignment can thus be seen as key factors of successful product development management [10], [19], [20]. In conclusion, high-performing enterprises should have the following key elements in place: 1) organisation-wide visibility on shared goals and their current progress, 2) real-time status tracking, and 3) comprehensive, flexible integration.

### B. Approaches to Achieving Organisational Alignment

GQM⁺Strategies is not the only approach that aims for organisational alignment. There are both general approaches and approaches that are specific to the software development domain. COBIT [21] is an approach for IT governance that distinguishes goals on different levels and provides a predefined set of metrics to measure the fulfilment of IT-specific goals. The predefined metrics offer some uniformity regardless of the application context. GQM⁺Strategies, on the other hand, supports customisation of measurement programs based on measurement goals. Practical Software and System Measurement (PSM) [22] is an approach to collect, analyse and manage aspects of software and software systems. Similarly to GQM⁺Strategies, software project characteristics and needs guide selection of suitable metrics to measure predefined software-specific goals. However, PSM mainly focuses on the project level and does not explicitly support goal measurement on higher organisational levels. The Quality Improvement Paradigm (QIP) [23] focuses on process and software quality improvement. QIP considers both on particular projects, but also feeds back into the organisation undertaking a software project. Through this feedback, organisations can align processes with their goals. QIP does not explicitly specify how goals on different levels of the organisation can be linked. However, QIP can be used in conjunction with both GQM+Strategies and the Goal-Question-Metric approach itself.

More general approaches to goal alignment exist outside of the software development domain. Balanced Scorecard (BSC) [24] can be used to align business activities and strategies of an organisation and to monitor organisational performance against set goals. BSC predefines four perspectives (financial, customer, internal business processes, learning and growth). Six Sigma [25] focuses on quality improvement through identification and elimination of defect causes in business and manufacturing processes. Unlike GQM+Strategies, BSC and Six Sigma do not explicitly support linking goals on different levels of an organisation in a rationale-driven way.

*C. Experiences Applying GQM+Strategies*

Several case studies on aligning goals and strategies with GQM+Strategies have been published. A search on the Scopus and Web of Science search engines using the keywords "gqm strategies case study" resulted in six articles. Two of these were theoretical comparisons between integrated measurement approaches [26], [27]. An empirical assessment of the practical value of GQM+Strategies was published by Mandic et al. [28]. This assessment used a revised version of the well-known Bloom's educational taxonomy to determine that for the company under examination, GQM+Strategies is valuable in practice and capable of addressing real-world problems related to goal-setting and goal alignment.

GQM+Strategies was also applied in a multi-organisational setting from the aerospace industry [29]. The approach helped to clarify and align the contributions of an internal organisational unit to the overall business-level goals of the entire organisation. Additionally, the measurement needs of projects with external suppliers were made transparent throughout the organisation.

The aforementioned experiences of using GQM+Strategies are not limited to "pure" software engineering – the approach has been applied in many different domains. Several other application cases have been reported, e.g, in the oil and gas industry and in the telecommunication sector. As such, the approach should be applicable in an IPD organisation.

## IV. RESEARCH METHODOLOGY AND EXECUTION

This section describes the target problem setting and the overall context of the company under study. The research approach and the implementation of the GQM+Strategies in this particular organisation are then explained.

*A. Context and Problem*

The organisational division under analysis is a new product line division in a larger, established company. The current main business driver of the unit is to get new market share by utilising latest technology with a strong network of component suppliers and partners.

TABLE I.   RESEARCH PROBLEM SETTING.

| Element | This Investigation |
|---|---|
| Analyse | Development process |
| In order to | Understand and improve |
| With respect to | Organisational visibility |
| From the perspective of | Platform Project Manager |
| In the context of | Distributed platform development |

The main offering is an integrated platform consisting of hardware and software components. The platform is a prefabricated product creation kit for device manufacturers to develop their new products rapidly. Such a platform accelerates development of customised devices and reduces development cost. The company supports this by providing the platform (hardware and software) coupled with customisation services. The service projects are out of the scope of this investigation. The current challenge is to manage platform development and customer projects by meeting schedule targets and commitments in predictable ways. By nature, there are mid-course changes in such development caused for example by hardware component modifications. The development process is already suited for such changes, since it is based on the Agile Scrum process model. Based on initial discussions with case company representatives, the main causes of schedule slippages are assumed to stem from the lack of visibility into the development status, making it difficult to foresee potential schedule problems early and in particular while making changes. While conventional project management techniques could be applied in specific projects to react to changes in external conditions, better goal alignment could reduce internal reasons for slippages.

The overall goal was thus to disentangle those current operations and information flows, and to make them aligned and transparent. In order to achieve those traits in the case company, we defined the initial problem statement in terms of the GQM model as shown in Table I.

*B. Research Approach*

The investigation was conducted collaboratively between the University of Helsinki research team and the industrial company representatives. The research team comprised four members: one principal scientist and a postdoc researcher coupled with one PhD-student researcher and one MSc-level student. The lead researcher possessed deep knowledge of and prior experience with the GQM+Strategies approach but it was new to the other members. The two former researchers had considerable industrial background including certain IPD organisations.

The industrial team included two senior R&D managers and two (senior) software development team members. The GQM+Strategies method was new for them also.

The starting point of the investigation was a set of preliminary goals outlined by the company key stakeholders. These goals were assumed to be the most significant ones contributing to the business success of the product development unit. However, no particular systematic process had been used to derive them, i.e., they were based mostly on intuition and prior experiences of the company. Because of the loosely-defined starting point and partial pre-understanding, we decided to engage with the company using mostly an interpretive research approach to begin with [30], [31]. Moreover, based on the preliminary information, we expected to face not only technical software engineering issues but also significant elements stemming from the social environment and organisational conditions in the company [10].

Following that line of preparations, we asked the company key representatives to provide the following kind of information for running joint workshops (see Table II):

- Organisation charts.
- Customer vs. platform project responsibilities.
- In-house terminology.
- Process descriptions.
- Existing measurement programs.
- Product and service descriptions.

### C. Execution

The application of GQM[+]Strategies consisted mainly of a set of planning meetings, followed by workshops with organisation representatives. During each workshop, key issues were discussed and goals were revised based on the latest information.

Following each workshop, a revised version of the in-progress GQM[+]Strategies model was developed by the research team and sent to the organisation representatives prior to the start of the next planned workshop. All workshop sessions except one were chaired by the same lead researcher. The sessions were recorded. Between the workshops there was limited occasional exchange of information with the industrial team.

Table II presents a timeline of the joint working events held during the application of GQM[+]Strategies, along with notable new information elicited during each event. The kick-off meeting, held in December 2011, focused primarily on understanding the objectives of the project. For the industrial organisation representatives, the application of GQM[+]Strategies was seen as a tool by which continuous planning and strategic measurement could be achieved. The overall objective, however, was increased transparency within the organisation. The actual meaning of this general concept would become clear once workshops were being held.

During the two subsequent planning meetings (held in January, 2012), general information was elicited, including information regarding the company domain and context, different project types (customer projects, platform projects), organisational structures and issues stemming from the geographic distribution of organisation offices and branches.

Being an IPD business, a key driver of software engineering was the development of a common, reusable hardware/software platform, from which specialised customer products are developed. Additionally, the loose structure of the organisational units and teams inside the organisation meant that a comprehensible view of structure and hierarchy would be challenging to elicit. A sound hierarchy is important in determining which goals belong to which organisational levels and thus also which goals could be potential sub-goals of a business-level goal.

Following the planning phase, the first project workshop was held in February 2012. First, a basic introduction of the GQM[+]Strategies approach was presented to the industrial team. During this workshop, initial goals for four different units of the organisation were elicited. The initial business goals were to improve predictability, schedule accuracy and transparency within the organisation. In addition to the insight that transparency meant increased visibility, a rough organisational hierarchy was established. The Business Unit (BU) is on the highest level, followed by Research and Development (R&D) and Project and Product Management (PPM). Work in discrete hardware and software development teams is overseen by the PPM unit and partially also the R&D branch.

The second workshop was held in April 2012 and focused on discussing the initial business-level goals in detail. The main hindrance to improved schedule accuracy was found to be the lack of proper cross-team dependency identification. According to the organisation representatives, identifying dependencies between development teams meant improved collaboration between the two – an understandable development-level strategy with obvious ties to improved visibility.

The discussion of schedule accuracy led to another insight: given accurate schedules, the overall predictability is improved, but the main reason to strive for improved predictability and visibility is to improve customer satisfaction in terms of keeping the delivery promises. This is a business-level goal that is implicit in all for-profit organisations – until now, this clear goal had not been included in the model-under-development. As a result of the second workshop, increased customer satisfaction was recognised as the foremost business-level goal. Predictability and transparency are factors that impact it. Improved product quality – a goal brought forward by both the PPM and R&D units – naturally contributes towards attaining customer satisfaction. Establishing a direct channel of communication between the development teams and product customers was suggested as a potential strategy for achieving such transparency.

The third workshop during our application of GQM[+]Strategies was held towards the end of April 2012. A better understanding of organisational structure was gained: the PPM and R&D units work on essentially the same hierarchical level, sharing common goals. When questioned how schedule adherence and accuracy may be improved, the representatives mentioned the continuous deployment of integrated hardware and software solutions as a possible target for improvement. It was noted that it is often the hardware development cycles that negatively affect adherence to schedules, but the software development alone is rarely the cause for product delays. By improving the hardware/software dependency identification through a higher degree of cross-team collaboration and communication, it was suggested that the continuous product deployment process would also improve. Increased collaboration was identified in the previous as a possible strategy also for increased organisational visibility. Overall, potential strategies for achieving the identified goals were discussed thoroughly for the first time during this workshop.

The results of the third workshop clearly showed that the iterative nature of GQM[+]Strategies stimulates organisations to identify apparent problem areas. Here, issues with predictability and scheduling stem from non-optimal hardware/software dependency identification and hardware development cycle accuracy, and can potentially be remedied via cross-team collaboration. How cross-team collaboration might be improved was also discussed in detail, with the suggestion of collaborative hardware/software teams with members from both domains generating the most interest – a well-known general strategy in IPD environments.

TABLE II. JOINT WORKING EVENTS, CORRESPONDING ELICITED INFORMATION, AND DISCOVERED CHALLENGES AND PROBLEMS.

| Timeline | Purpose | Elicited Information | Challenges and Problems |
|---|---|---|---|
| Start (December 2011) | Research project kick-off meeting | Project objectives | *How should the aims and benefits of GQM⁺Strategies be communicated effectively?* |
| Approx. one month after start (15.1.2012) | Familiarisation with company objectives, domain, context and underlying assumptions | • Company business and technology context<br>• Project types (customer and platform)<br>• Organisational structures<br>• Cultural issues (e.g., multisite differences) | *What information is relevant when applying GQM⁺Strategies?* |
| Approx. six weeks after start (27.1.2012) | Familiarisation with company objectives, domain, context and underlying assumptions | • Platform characteristics (software+hardware), product areas<br>• Company-specific Scrum process<br>• Cultural aspects<br>• Potential measurements | *How should one measure the value of the platform development vs. customer projects?* |
| Approx. two months after start (15.2.2012, general workshop) | Familiarisation with company objectives, domain, context and underlying assumptions | The target organisation is looking for a light-weight way of applying GQM⁺Strategies in an Agile/Lean enterprise | *What types of strategies can be implemented in a light-weight fashion?* |
| Approx. nine weeks after start (21.2.2012, project workshop 1) | Initial GQM⁺Strategies model development | Initial set of elicited goals:<br>• Improve predictability<br>• Improve schedule accuracy<br>• Improve organisational visibility and transparency | *What are the main differences between organisational visibility and transparency? How should the GQM⁺Strategies models be presented to target organisations?* |
| Approx. 16 weeks after start (10.4.2012, project workshop 2) | Refinement of the GQM⁺Strategies model | • Cross-team dependency identification key to accurate execution of product plans<br>• Improved cross-team visibility and collaboration key to improved dependency identification<br>• Accurate scheduling key to increasing predictability<br>• Predictability recognised as an important factor in delivering on commitments made to customers<br>• Recognition of improved customer satisfaction as the most important business goal<br>• Increased customer satisfaction entails better adherence to schedules and increased product quality<br>• Product quality is likely to increase if development teams communicate directly with customers | *What are key factors for improved customer satisfaction in the domain of integrated hardware / software development?* |
| Approx. four months after start (20.4.2012, project workshop 3) | Refinement of the GQM⁺Strategies model | • More detailed information on organisational structure<br>• Success of continuous deployment key to accurate execution of product plans and thus also adherence to schedules<br>• Improving the hardware/software integration process leads to improved continuous deployment<br>• Hardware/software dependency management key to improved hardware/software integration<br>• Improved hardware/software dependency identification requires better visibility between hardware/software teams | *How should the goals of a flat organisation be represented in GQM⁺Strategies? What are key metrics for hardware/software integration and continuous deployment?* |
| Approx. 20 weeks after start (3.5.2012, project workshop 4) | Refinement of the GQM⁺Strategies model | • The overall strategic intent is to improve the lead-time for the platform product customers<br>• High utilisation of the organisation's hardware/software platform can help reduce delays during product development<br>• Some information regarding data that is being collected by the organisation and could be used when developing GQM goals and metrics | *How should one measure the cost and effort of maintaining a common hardware / software platform?* |
| Approx. five months after start (22.5.2012, project workshop 5) | Refinement of the GQM⁺Strategies model | No significant new information | *When should a GQM⁺Strategies model be considered complete?* |

In preparation for the fourth workshop, details of the targeted common hardware and software product platform were revisited, and potential goals to improve its utilisation and thus increase schedule accuracy were defined. This required significant rework of both the R&D and PPM goals, but was well received during the workshop, with only minor changes requested. At this stage, the GQM+Strategies model-under-development was quite close to the last model revision. Notably, this revision included a set of fully developed strategies, such as the introduction of direct communication between developers and customers and possible establishment of an integrated hardware/software team. Overall, the primary focus of the fourth workshop was discussing what measurement data was already being collected by the organisation. The aim was to identify what historical information could be of use when further developing the GQM goals and their associated strategy metrics.

The fifth and final workshop during the application of the GQM+Strategies approach suggested only minor changes to the model-under-development, suggesting that the approach has resulted in the successful identification and alignment of key organisational goals.

To summarise, the main concerns discovered during the workshops were ones related to organisational distribution, cross-disciplinary engineering, and product structure.

## V. Lessons Learned

During the goal elicitation process, we identified a number of issues that, if corrected, would improve our application of the GQM+Strategies approach in subsequent business cases. The first such issue pertains to knowledge of GQM+Strategies itself: for representatives of the target organisation and some researchers, there was no prior experience in applying the approach. Some initial training on the process itself would likely have increased organisation representatives' understanding of what information is significant and thus provided more structure to workshops.

One of the challenges in applying GQM+Strategies is finding suitable entry points, i.e., to carefully select on which level, with which stakeholders, and with which goals to start. In this case, many goals were deemed too abstract at the beginning of the goal elicitation phase to warrant a detailed definition of associated GQM graphs. In addition, we have seen in this case that it was helpful to elicit goals on different levels first and not to integrate the strategies in the beginning. The distinction between goals and strategies was seen as difficult in the early stages and we recommend to include strategies in a later phase in the process.

Another challenge is understanding the organisational structure, especially with respect to roles. In this case, we found roles that were not clearly defined, which made it difficult to establish facts and assumptions regarding those roles for the GQM+Strategies grid, and to understand information needs related to them. A number of reasons may explain lack of clarity in roles, e.g., a recent or ongoing organisational change, a company culture that assumes high flexibility in individual roles, or even a brittle organisational architecture. In this case, the lack of clarity in some roles were found to stem from recent changes, which meant that the roles were not yet fully understood. Regardless of the reason, the role-related assumptions should be carefully examined. We found it very useful to have company participants from multiple levels in the organisation. This allowed us to pinpoint role-related assumptions in different parts of the organisation. Such roles could be made clearer by combining the understandings and devising a clarifying GQM+Strategies grid.

During the later stages of the process, we found that a flexible tool for making iterative changes to the grid would have been very helpful. The tools we used included spreadsheets and graphics programs, but these were too cumbersome to use in real time during workshops. A tool which operates on the basic building blocks of the GQM+Strategies approach (goals, strategies, facts, assumptions, measures) would have allowed us to record our understanding of the information given by company representatives immediately. A visual representation of the grid would have allowed the company representatives to give immediate feedback and the process could have resulted in deeper insights in shorter time. However, this would have been useful mainly during the later stages of the process, when a significant portion of the material had been gathered and large portions of the grid were in place. At the start of the process, we found that so many open questions regarding the case company needed to be examined that a tool would likely not have helped significantly.

Finally, a major factor that was repeatedly encountered was the amount of change anticipated by the company representatives. A significant amount of effort was spent on designing goals and strategies that were convincing and credible in practice. Some goals and strategies were discarded because they were likely to be obsolete after a short period of time. While we did not discover any generalisable method by which a GQM+Strategies grid could be checked for flexibility under changing circumstances, we found that utilising experience from domain experts was crucial. Also, we found it more fruitful to propose several partial alternatives and adjust them based on feedback than attempting to propose large parts of a grid at once. The criterion of saturation, common in case study and qualitative research, can be applied here: the grid tends to stabilise over time.

Overall, the method significantly helped to clarify and harmonise the key goals across the organisation. Many new links between different goals could be identified and a multitude of activities could be linked in a way that they clearly contribute to higher-level goals. Systematically gaining and using experience on how to apply GQM+Strategies in organisations is needed and will help to guide future alignment activities in organisations.

Due to the iterative nature of GQM+Strategies, goals may evolve in a variety of ways. By formalising all possible goal evolution patterns or transformations, it is possible to identify a set of explanations as to why each one occurs and under what circumstances. By identifying these circumstances, a set of criteria for when to apply a particular transformation could possibly be devised. Such criteria can serve as a handbook for practitioners of GQM+Strategies, helping to determine when to apply a transformation and thus reach a sound, balanced model in a shorter period of time.

Table III presents some example goal evolution patterns along with plausible explanations for their occurrence. An expanded, formal definition of all possible patterns and prerequisites for their application is outside the scope of this paper and thus reserved for future work.

TABLE III.  EXAMPLE GOAL EVOLUTION PATTERNS.

| Pattern | Plausible Explanations |
|---|---|
| Unchanged throughout the process | The goal was really well-set and understood from the beginning. The original (high-level) goal was later taken into account with some lower-level goal(s) (possibly implicitly). |
| Revised during the process | The goal remained important but was originally defined in an ambiguous fashion. |
| Discarded during the process | The goal was not really important after all. The original (high-level) goal was incorporated into another (lower-level) goal. |
| Introduced during the process | Previously unknown information led to a new goal being defined. Previously elicited information contained a clear goal, but went unnoticed. |
| Introduced from splitting another goal during the process | The original goal was a combination of goals, and should logically be split into two or more distinct goals. |
| Merged into another goal during the process | The original goal was so narrow in scope or too granular that a single GQM graph for all similar and similarly granular goals could be devised. The same goal is defined for another unit on the same organisational level. |

## VI. CONCLUSIONS AND FUTURE WORK

In this paper, we have presented a case of applying the GQM⁺Strategies method to examine and align the strategic, tactical, and operational goals in software-intensive integrated product development. We observed the following key benefits and advantages:

- The methodology brought systems thinking into this systems development process as well as in the engineering and organisational design. This is intrinsically important, given that not only the products themselves but also the product development organisation is a large complex system.
- The GQM⁺Strategies method offered a common, conceptual framing for the discussion and analysis. It has rich concepts and is a theoretically sound approach and yet readily comprehensible and actionable in practical settings.
- The iterative way of working directed the goal refinement in a flexible manner. This is an advantage in particular when initial goals are uncertain and subject to even significant revising.

One of the key lessons from our application GQM⁺Strategies is that goal discovery is hardly a straightforward process in practice. In particular, the initial top-level goals may not be the ultimate key goals after all, and some of the early high-priority goals may turn out to be less significant. Consequently, the emerging goal-set should be traversed in different ways, i.e., bottom-up, top-down and horizontally.

For example, this case discovered an underlying goal that was not initially anticipated. The most prevalent top-level goal in the beginning of the study centred around the theme of transparency. During the process, we discovered that this goal was actually closely related to managing customer satisfaction. Transparency was a sub-goal that had been implicitly identified as impacting this higher-level goal. Thus the application of GQM⁺Strategies can help prioritise what activities are really important for a company in order to achieve their strategic business goals. We recommend to repeatedly and constructively question the priority order of the goals as well as to find out whether there are actually hidden, higher-level goals that have not yet been discovered.

Another major learning was that goals may not be limited to basic strategic, tactical, and operational goals. In addition, softer elements of an organisation – motivational factors, different traditions in different engineering disciplines and various cultural constraints – should be taken into consideration, since these may have significant impact when implementing associated strategies.

Our planned future work entails continued development and implementation of strategies to achieve the aligned customer satisfaction goals in practice for the company under study. In addition, metrics to determine the achievement of each goal are to be developed.

Moreover, we plan to investigate further the prospect of devising a set of generalised goal-evolution patterns like illustrated in Table III. That would lead us to new potential improvement measures also in other organisations.

ACKNOWLEDGEMENTS

This work was supported by Cloud Software Program which is funded by TIVIT (Finnish Strategic Centre for Science, Technology and Innovation in the field of ICT) and TEKES.

REFERENCES

[1] J. Pfeffer, "Seven practices of successful organizations," *California Management Review*, vol. 40, no. 2, pp. 96–124, 1998.
[2] M. Staron, W. Meding, G. Karlsson, and C. Nilsson, "Developing measurement systems: an industrial case study," *Journal of Software Maintenance and Evolution: Research and Practice*, vol. 23, no. 2, pp. 89–107, 2011.
[3] V. R. Basili, M. Lindvall, M. Regardie, C. Seaman, J. Heidrich, J. Munch, D. Rombach, and A. Trendowicz, "Linking software development and business strategy through measurement," *Computer*, vol. 43, no. 4, pp. 57–65, 2010.
[4] G. Caldiera, V. R. Basili, and H. D. Rombach, "The goal question metric approach," *Encyclopedia of software engineering*, vol. 2, pp. 528–532, 1994.
[5] V. Mandić, V. Basili, L. Harjumaa, M. Oivo, and J. Markkula, "Utilizing GQM+ Strategies for business value analysis: An approach for evaluating business goals," in *Proceedings of the 2010 ACM-IEEE International Symposium on Empirical Software Engineering and Measurement*. ACM, 2010, p. 20.
[6] M. B. Lyne, "Aligning R&D with Business Strategy," *Research-Technology Management*, vol. 46, no. 6, pp. 44–46, 2003.
[7] J. Peppard, "Bridging the gap between the IS organization and the rest of the business: plotting a route," *Information Systems Journal*, vol. 11, no. 3, pp. 249–270, 2001.
[8] K. Rautiainen, C. Lassenius, J. Nihtilä, and R. Sulonen, "Key Issues in New Product Development Controllability Improvement – Lessons Learned from European High-Tech Industries," in *Proceedings of the Portland International Conference on Management of Engineering and Technology (PICMET'99)*, 1999.
[9] V. Krishnan and K. T. Ulrich, "Product development decisions: A review of the literature," *Management Science*, vol. 47, no. 1, pp. 1–21, 2001.

[10] B. Curtis, H. Krasner, and N. Iscoe, "A field study of the software design process for large systems," *Communications of the ACM*, vol. 31, no. 11, pp. 1268–1287, 1988.

[11] A. Dagnino, "Coordination of hardware manufacturing and software development lifecycles for integrated systems development," in *Systems, Man, and Cybernetics, 2001 IEEE International Conference on*, vol. 3. IEEE, 2001, pp. 1850–1855.

[12] O. Jaufman and S. Przewoznik, "Suitability of state of the art methods for interdisciplinary system development in automotive industry," in *Proceedings of the 2004 ACM workshop on Interdisciplinary software engineering research*. ACM, 2004, pp. 78–82.

[13] D. Karlstrom and P. Runeson, "Combining agile methods with stage-gate project management," *Software, IEEE*, vol. 22, no. 3, pp. 43–49, 2005.

[14] A. Genus and A.-M. Coles, "Firm strategies for risk management in innovation," *International Journal of Innovation Management*, vol. 10, no. 02, pp. 113–126, 2006.

[15] R. Dove, "Fundamental principles for agile systems engineering," in *Conference on Systems Engineering Research (CSER), Stevens Institute of Technology, Hoboken, NJ*, 2005.

[16] P. Patanakul and A. Shenhar, "Exploring the concept of value creation in program planning and systems engineering processes," *Systems Engineering*, vol. 13, no. 4, pp. 340–352, 2010.

[17] R. Turner, "Toward Agile systems engineering processes," *Crosstalk. The Journal of Defense Software Engineering*, pp. 11–15, 2007.

[18] V. Allee, "Value network analysis and value conversion of tangible and intangible assets," *Journal of Intellectual Capital*, vol. 9, no. 1, pp. 5–24, 2008.

[19] M. McGrath, *Next Generation Product Development: How to Increase Productivity, Cut Costs, and Reduce Cycle Times*. McGraw-Hill Education, 2004.

[20] H. Sicotte and A. Langley, "Integration mechanisms and R&D project performance," *Journal of Engineering and Technology Management*, vol. 17, no. 1, pp. 1–37, 2000.

[21] J. W. Lainhart IV, "COBIT$^{TM}$: A methodology for managing and controlling information and information technology risks and vulnerabilities," *Journal of Information Systems*, vol. 14, no. s-1, pp. 21–25, 2000.

[22] J. McGarry, D. Card, C. Jones, B. Layman, E. Clark, J. Dean, and F. Hall, *Practical Software Measurement: Objective Information for Decision Makers*, 1st ed. Addison-Wesley Professional, 2001.

[23] V. R. Basili, G. Caldiera, and H. D. Rombach, "Experience factory," *Encyclopedia of software engineering*, 1994.

[24] R. S. Kaplan, D. P. Norton *et al.*, "The balanced scorecard–measures that drive performance," *Harvard business review*, vol. 70, no. 1, pp. 71–79, 1992.

[25] Z. Pan, H. Park, J. Baik, and H. Choi, "A Six Sigma framework for software process improvements and its implementation," in *14th Asia-Pacific Software Engineering Conference, 2007. APSEC 2007.* IEEE, 2007, pp. 446–453.

[26] L. Olsina, F. Papa, and P. Becker, "Assessing integrated measurement and evaluation strategies: A case study," Moscow, 2011.

[27] F. Papa, "Toward the improvement of a measurement and evaluation strategy from a comparative study," *Lecture Notes in Computer Science (including subseries Lecture Notes in Artificial Intelligence and Lecture Notes in Bioinformatics)*, vol. 7703 LNCS, pp. 189–203, 2012.

[28] V. Mandić, L. Harjumaa, J. Markkula, and M. Oivo, "Early empirical assessment of the practical value of GQM$^+$Strategies," *New Modeling Concepts for Today's Software Processes*, pp. 14–25, 2010.

[29] M. Kowalczyk, J. Münch, M. Katahira, T. Kaneko, Y. Miyamoto, and Y. Koishi, "Aligning Software-related Strategies in Multi-Organizational Settings," in *Proceedings of the International Conference on Software Process and Product Measurement (IWSM/MetriKon/Mensura 2010), Stuttgart, Germany*, 2010.

[30] A. C. Edmondson and S. E. McManus, "Methodological Fit in Management Field Research," *Academy of Management Review*, vol. 32, no. 4, pp. 1155–1179, 2007.

[31] E. Gummesson, *Qualitative methods in management research*. Sage Publications, Incorporated, 1999.